\documentclass[a4paper]{jpconf}
\usepackage{graphicx}
\graphicspath{{figs/}}

\begin{document}
\title{Noncommutative Black Holes at the LHC}

\author{Elena Michelle Villhauer}

\address{University of California, Irvine; ATLAS, CERN}

\ead{elena.michelle.villhauer@cern.ch}

\begin{abstract}
Based on the latest public results, 13 TeV data from the Large Hadron Collider at CERN has not indicated any evidence of hitherto tested models of quantum black holes, semiclassical black holes, or string balls. Such models have predicted signatures of particles with high transverse momenta. Noncommutative black holes remain an untested model of TeV-scale gravity that offers the starkly different signature of particles with relatively low transverse momenta. Considerations for a search for charged noncommutative black holes using the ATLAS detector will be discussed. 
\end{abstract}

\section{Introduction to the Hierarchy Problem and Black Hole Formation at the LHC}

The European Organization of Nuclear Research, CERN, seeks to study the Standard Model of particle physics, its problems, and possible solutions. Arguably the most outstanding and pressing problem that the Standard Model fails to address is the irreconcilability of general relativity and quantum mechanics.  Quantum gravity is a theory that attempts to explain gravity according to the principles of quantum mechanics.  The quantum effects of gravity are believed to become strong and apparent at the Planck scale, which, in four dimensions, is estimated to be around 1 x $10^{19}$ GeV. Another prominent and pressing problem that the Standard Model fails to address is the hierarchy problem, or the extreme discrepancy between aspects of gravity and the weak force. Out of the four fundamental forces, the gravitational force is by far the weakest, ranking approximately 32 orders of magnitude weaker than the weak force, 38 orders of magnitude weaker than the strong force, and 36 orders of magnitude weaker than the electromagnetic force. As noted by Arkani-Hamed, S. Dimopoulos and G. Dvali, one way to account for the extreme discrepancy in gravity's relative strength, the hierarchy problem, is the existence of extra dimensions~\cite{ADDpheno, ADD}. Gravity, unlike the other fundamental forces, may propagate in more than four dimensions, thereby reducing its apparent strength. With more than four dimensions, the Planck scale may be lowered to the low TeV-scale, the energy range accessible by the Large Hadron Collider in Geneva, Switzerland~\cite{ADD}. This allows for the possibility of black hole production. A black hole is predicted to be formed at the LHC if, as proposed by Thorne, the hoop conjecture is fulfilled by two partons from a proton-proton collision~\cite{Casadio, Thorne}. The classical Hoop Conjecture states that a black hole can be formed if an imploding object is compressed in all directions into a region with a circumference smaller than 2$\pi$ times the object's Schwarzschild radius~\cite{Thorne}. While LHC black hole event generators incorporate the classical hoop conjecture, a more precise and accurate description of the hoop conjecture for quantum black hole production at the LHC can be found in a study of the quantum hoop conjecture by Casadio, Micu, and Scardigli~\cite{Casadio}. The main types of black hole and signatures of TeV-scale predicted to be formed at the LHC are semiclassical black holes, quantum black holes, and string balls. The latest results for searches for each type of black hole formation will follow. 

\section{Models Tested and 13 TeV Results}

\subsection{Semiclassical Black Holes}

Semiclassical black holes at the LHC are primarily modeled by ATLAS and CMS with the Charybdis2 and BlackMax generators using ADD-type extra dimensions. Two ATLAS semiclassical black hole searches have been performed with 13 TeV data: a multijet search and a lepton plus jets search. For the multijet search, the selection included at least three jets with scalar sum of jet transverse momenta greater than 1 TeV. Using 13 TeV data with an integrated luminosity of 3.6 fb$^{-1}$, the ATLAS multijet search set exclusion limits for rotating black holes in 6 extra dimensions with mimumum black hole masses of 9.0 TeV - 9.7 TeV~\cite{ATLASmultijet}. The lepton plus jet search set a selection consisting of at least three objects: a leading lepton with pT $>$ 100 GeV and at least two other objects (jets or leptons) with pT $>$ 100 GeV. Additionally, the $\sum$ pT was required to be $>$ 2 TeV or 3 TeV. The lepton plus jet search used data with an integrated luminosity of 3.2 fb$^{-1}$ to exclude rotating black holes in two extra dimensions with minimum black hole masses up to 7.8 TeV for MD = 2 TeV and with minimum black hole masses of 7.4 TeV for MD = 5 TeV~\cite{test}. CMS undertook a search for semiclassical black holes decaying to multijets, in which data of an integrated luminosity of 2.3 fb$^{-1}$ was used to exclude semiclassical black holes with masses as high as 9.5 TeV~\cite{CMSMultijets}.

\subsection{Quantum Black Holes}


The ATLAS experiment has set limits on the following QBH decay channels using 13 TeV data: QBH $\rightarrow$ dijet, QBH $\rightarrow$ photon + jet. The ATLAS dijet search used data with 37 fb$^{-1}$ to set a 95 percent confidence level exclusion on quantum black holes up to masses of 8.9 TeV~\cite{ATLASdijet}. The ATLAS photon + jet search used data with an integrated luminosity of 36.7 fb$^{-1}$ to exclude QBH in RS-type extra dimensions below 4.4 TeV and QBH in ADD-type extra dimensions below 7.1 TeV~\cite{ATLASphoton}. The CMS experiment has used 13 TeV data with an integrated luminosity of 3 fb$^{-1}$ to set limits on QBH $\rightarrow$ dijet for 6 dimensional ADD-type QBH with a lower limit mass of 7.8 TeV and for 5 dimensional RS-type QBH with a lower limit mass of 5.3 TeV. The CMS multijet used 13 TeV data with an integrated luminosity of 2.3 fb$^{-1}$ to exclude quantum black hole masses as high as 9.0 TeV~\cite{CMSMultijets}.

\subsection{String Balls}

 
ATLAS searches for stringballs in the multijet final state~\cite{ATLASmultijet} using data 3.0 fb$^{-1}$ to exclude string balls as high as 9 TeV. CMS used data of an integrated luminosity of 2.3 fb$^{-1}$ to exclude string balls with masses as high as 9.5 TeV~\cite{CMSMultijets}.

\section{Results Interpretation}
\subsection{Reasons why extra dimensions have not yet been discovered}
ATLAS and CMS have used 13 TV data from the 2015-2017 Run II campaign to search for semiclassical black holes, quantum black holes, and string balls. Based on the latest public results, no evidence of such models has been found and exclusion limits have been set. The absence of evidence for signatures of TeV-scale gravity at the LHC can be attributed to either one of three reasons. It could be that extra dimensions do not exist and any subsequent search for signatures of extra dimensions will prove futile. However, it could also be that extra dimensions do exist, and that we have not discovered evidence of them because we need more energy to access them. Finally, and most excitingly, signatures of TeV-scale gravity may have already been produced at the LHC, but have either been hiding in signatures different than those we have used in searches or because they have been produced at a very low rate and the signal is diluted.

\subsection{The LHC's future plans and the best bet for TeV-scale gravity}

The LHC has been producing collisions at 13 TeV since April 2015. However, at the end of 2018, the LHC plans to enter a long shutdown, at which point all data taking will stop. Collisions are scheduled to resume at the start of 2020 with energies of 14 TeV. This is the only planned energy increase in the set future of the LHC. All other improvements scheduled until 2035 will be regarding luminosity, as part of the High Luminosity Large Hadron Collider(the HL-LHC), the version of the LHC intended for luminosity upgrades up to 3000 fb$^{-1}$~\cite{HLLHC}. This poses a problem for models of TeV-scale gravity already tested because black hole cross sections increase exponentially with increases in energy, but only linearly with increases in luminosity. Upon increasing the energy of the LHC from 8 TeV to 13 TeV, the 13 TeV to 8 TeV cross section ratio of quantum black holes with masses of 6 TeV becomes 9000. When increasing from 13 TeV to 14 TeV, the 14 TeV to 13 TeV cross section ratio of quantum black holes with masses of 9 TeV in 6 extra dimensions becomes 5.3 TeV.  Larger increases in energy, therefore, would provide the best opportunity for discovery. However, increases in energy will not occur until after the long shutdown at the winter of 2018, after which energy is set to increase to 14 TeV. There is perhaps a possibility to increase energies by replacing a magnet. However, no such plans have been set for the near future. Therefore, the best bet for finding new physics with black hole remains looking at different models yielding different signatures. Searches for semiclassical black holes, quantum black holes, and other signatures of TeV-scale gravity at the LHC have all targeted final states containing high transverse momenta. Noncommutative black holes offer a very different experimental signature from other potential signatures of TeV-scale gravity: many particles with a low transverse momenta spectra.

\section{Noncommutative Black Holes}

As described by Nicolini~\cite{Nicolini}, noncommutative black holes are predicted to form in models where noncommutative geometry is incorporated into the framework of extra dimensions. The defining feature of noncommutativity in black holes is that the spacetime coordinates $\hat{x}^A$ and $\hat{x}^B$ do not commute~\cite{Nicolini}, such that
\begin{equation} \label{eq1}
\left[ \hat{x}^A, \hat{x}^B \right] = i\theta^{AB} \equiv i
\frac{\epsilon^{AB}}{\Lambda_\mathrm{NC}^2}\, ,  
\end{equation}

\noindent
where, taking directly from Gingrich~\cite{DGNC}, ``$\theta^{AB}$ is an real antisymmetric $D\times D$ matrix.
For convenience, we have separated the mass scale $\Lambda_\mathrm{NC}$
associated with noncommutative from the dimensionless matrix structure
$\epsilon^{AB}$ of $\theta^{AB}$.
If $\Lambda_\mathrm{NC}^{-2}$ is the average magnitude of the elements
in $\theta^{AB}$, we assume the elements of $\epsilon^{AB}$ are of 
$\mathcal{O}(1)$.''

Noncommutativity invokes a smearing in the mass distribution~\cite{FNCBH} such that the charge density~\cite{TRNC}, $\rho$, becomes~\cite{DGNC} 
\begin{equation} \label{eq2}
\rho = \frac{m}{(4\pi\theta)^{(n+3)/2}} e^{-r^2/(4\theta)}\, ,
\end{equation}

\noindent
where $\sqrt{\theta} = 1/\Lambda_\mathrm{NC}$

Noncommutative black holes are theoretically of interest due to the resulting nonsingular solution, but most importantly, because they form an effective theory to quantum gravity. Experimentally, this effective theory allows for a new mass threshold above the Planck scale and, thus, a new energy regime for physics beyond the Standard Model~\cite{DGNC}. Noncommutative black holes are cold and decay to a stable remnant~\cite{FNCBH},~\cite{Nicolini}. The opportunity to study stable remnants through noncommutative black hole studies is exciting because stable remnants appear in a variety of models ranging from those of loop quantum gravity and string gravity to tunneling~\cite{DGNC}. 

\section{2010 Neutral Noncommutative Black Hole Study}

A study undertaken in 2010 by Gingrich~\cite{DGNC} shows the phenomenological aspects of noncommutative black holes at the LHC. The cross sections and temperatures for noncommutative black holes, shown in Figure~2~\cite{DGNC}, are significantly lower than those of other signatures of TeV-scale gravity. The combined effect is that noncommutative black holes yield softer particles than semiclassical black holes, string balls, or quantum black holes.

\begin{figure}[h]
    \centering
    \includegraphics[width=\textwidth]{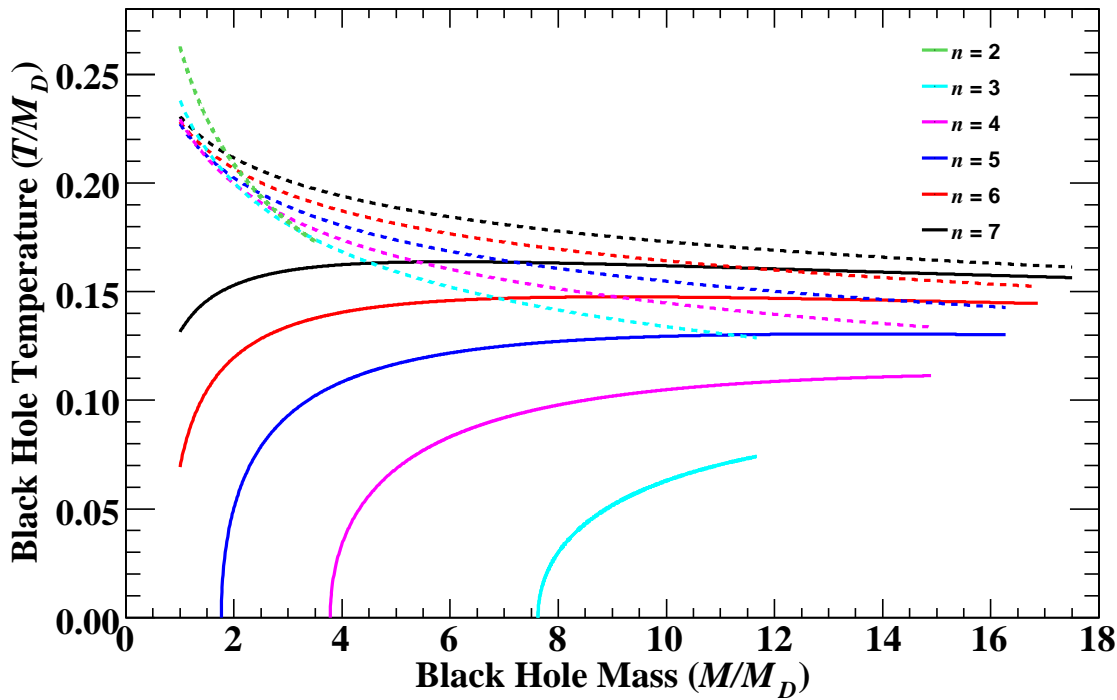}
    \caption{\label{label} Black Hole Temperature vs. Black Hole Mass (M/$M_{D}$). Solid lines: NC BH, Dashed lines: Commutative BH. Figure taken from~\cite{DGNC}.}
\end{figure}

There is an average of 9 primary decay particles from the black hole and a maximum of 25. The average transverse momentum of the final state particles is 70 GeV. A generator cut-off was imposed at 100 MeV above the mass of the remnant due to technical complications regarding the generator's efficiency near the mass of the remnant~\cite{DGNC}. Without this cutoff, the primary decay particle multiplicity would be greater and the average transverse momentum of final state particles would be even lower. The majority of events in the simulations of the study contain at least one jet, while about 45 percent have a leading lepton. Figure 2~\cite{DGNC} depicts the soft jet maximum transverse momentum and the soft lepton maximum transverse momenta. This soft transverse momentum spectrum of the final state particles allows for the noncommutative black hole signal to blend into QCD background. Searches for semiclassical black holes with low masses set a selection on scalar sum of transverse momentum of greater than a few TeV. Due to the cold nature of noncommutative black holes, the scalar sum of transverse momentum can reach zero. This eliminates the ability to include scalar sum of transverse momentum in the selection~\cite{DGNC}.

As for the characteristics of the noncommutative black hole's stable remnant, the remnant is slow, with the most likely speed being .3c, and has an average transverse momentum of less than 230 GeV.

\begin{figure}[h]
    \centering
    \includegraphics[width=0.9\textwidth]{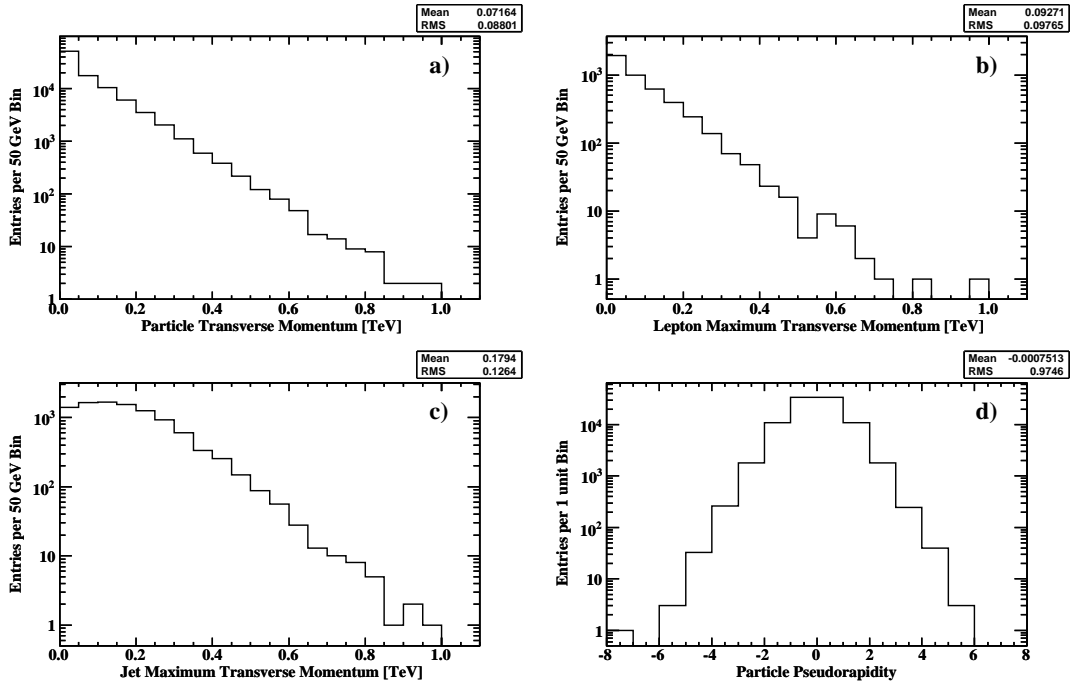}
    \caption{\label{label} Characteristics of Final State Particles. Figure taken from~\cite{DGNC}.}
\end{figure}

\section{Strategies for a Charged Noncommutative Black Hole Search}

The 2010 LHC noncommutative black hole studies by Gingrich~\cite{DGNC} show that noncommutative black holes offer the unique signature of particles with low transverse momentum. This was the only study of noncommutative black holes using LHC data, and it was stopped at trigger level due to the low transverse momentum spectrum of the final state particles being obscured by the QCD background. This study was undertaken for a neutral noncommutative black hole model. The addition of charge would further lower the transverse momentum spectrum, leading to a more dramatic signature. Therefore, a search for charged noncommutative black holes, in which the charge of the black hole is confined to the brane, is intended. The following is a discussion regarding possible search options and ways to separate the low pT signal from the QCD background.

A first step would be to look at the HT for each noncommutative black hole event and compare it to that of QCD. Thorough studies should be undertaken to compare the shape of black hole radiation to that of QCD. It might be useful to check if the black hole thermal distribution matches the KNO distribution. The signal may be able to be extracted from the QCD background by treating the signal and background as multijets and studying the isotropicness of the noncommutative black hole versus the isotropicness of the QCD background.

Adding a W or Z to the selection could help reduce the background, but comes at the risk of chopping the probability by a factor of alpha. For charged noncommutative black holes, adding a photon in the selection could help separate the signal from background. A search for multijets plus photons has not been done before, so this would have the added benefit of exploring a new final state.

The study by Gingrich points out that another way to detect noncommutative black holes is to search for the remnant. If the remnant is charged, once can use the fact that the remnant is slow to search for charged, long lived particles. Additionally, for a charged remnant, there should be a balancing charge. Perhaps this balancing charge looks like a charged lepton combiation, which can be used as a selection requirement.

\section{Conclusion}
Based on the latest public results, no evidence of hitherto tested models of quantum black holes, semiclassical black holes, or string balls have been found at the LHC using 13 TeV data. Noncommutative black holes offer a chance to use a different signature to search for TeV-scale gravity. A search is intended for charged noncommutative black holes using the ATLAS detector

\section*{Acknowledgments}

The author wishes to thank Marco Sampaio for the crucial specification that, for a charged black hole model to be incorporated in Charybdis2, the charge of the black hole must be confined to the brane. Additionally, the author would like to thank Frank Krauss, Richard Keith Ellis, and Daniel Maitre for their keen insight on separating low pT signal from QCD background, which was subsequently included above.


\end{document}